\documentclass[12pt,aps,nofootinbib,showkeys]{revtex4}
\usepackage{color}
\usepackage{multirow}
\usepackage{amsmath}
\usepackage{esvect}

\usepackage{amsfonts}    
\usepackage{amssymb}
\usepackage{latexsym}
\usepackage[toc,page]{appendix}
\usepackage{graphicx}
\usepackage[english]{babel}

\begin{document} 


\title{Manifestation of instability in the  quasiclassical limit of the spectrum of the quartic double well
} 





\author{D~.~J.~ Nader}
 \email{daniel_nader@brown.edu} 
 \affiliation{Department of Chemistry, Brown University, Providence, Rhode Island 02912, United States}
 
\author{J.~R.~ Hern\'andez-Gonz\'alez}
\affiliation{Facultad de F\'isica, Universidad Veracruzana, C. P. 91097, Xalapa, M\'exico.}

\author{H.~V\'azquez-S\'anchez}
\affiliation{Facultad de F\'isica, Universidad Veracruzana, C. P. 91097, Xalapa, M\'exico.}

\author{S.~Lerma-Hern\'andez}
 \email{slerma@uv.mx} 
\affiliation{Facultad de F\'isica, Universidad Veracruzana, C. P. 91097, Xalapa, M\'exico.}



\begin{abstract}

Within the Bohr’s correspondence principle, the quantum theory should
reproduce the classical world when   $\hbar \to0$. In practice, the discrete energies come close to each other before passing to continuum, causing that highly excited states become inaccessible. However it is possible to identify signatures that the spectrum is approaching the continuum limit. In this work, for $\hbar \to 0$,   we focus 
our attention on the spectrum of the 1D quartic double well and find signatures of the classical instability in the quantum spectrum near the critical
energy, and  explore how the tunneling phenomenon remains only at energies close to this critical energy.

\end{abstract}


\maketitle

\section{\label{sec:intro}Introduction}

Since the beginning of quantum mechanics, potentials which admit analytical solution
have played a key role to provide insights of the phenomena at the quantum scale. 
Right after them, we have non-analytically solvable problems 
which can be only accessed via approximation methods. One of the simplest non-analytically 
solvable problems is the quartic oscillator including both quadratic and 
quartic terms \cite{Lai,REID1970183,doi:10.1063/1.531962,Jing,Turbiner_delvalle,Turbiner_book}. In particular, for negative quadratic term added to a positive quartic term, we have the 
double well with a Mexican hat symmetric form which provides one of the  simplest examples 
of symmetry breaking \cite{Joger,DELABAERE1997180,meier}. The double well problem offers a good model to test 
theories of tunneling and to explore asymptotic approximations \cite{fujimura,Joger,meier}.
The ground and lowest states are well established and can be reached with a desired accuracy \cite{Fernandez2017}
by several approximation methods even though some of them show faster 
convergence than others (see discussion \cite{Okun,delvalle_comment} for instance). 
Less studied is the regime of highly excited states in the spectrum.

By its simplicity, the quartic potential also offers a good chance 
to explore quantum-classical correspondence \cite{Plata_1992}. For example, in \cite{Novaes_2003} a correspondence between 
the Husimi function and the classical trajectory for high quantum numbers was found and it was concluded that the zero point energy 
is relevant when dealing with low energy states. Similar correspondences have been observed for many-body models between the trajectory of the semiclassical approximation and the Husimi functions \cite{Aguiar_1991,FURUYA1992313,NaderPRE2021,Arranz1,PhysRevA.88.043835}.
Apart from the Husimi and trajectories, classical correspondence can also be observed in the density of states \cite{PhysRevA.89.032101,STRANSKY201473,LUNAACOSTA2000192,Puebla}. However this can be achieved only if highly excited states are estimated to a certain accuracy. 

Comparing quantum and classical analysis is specially valuable if some signatures of chaotic behavior can persist in the quantum analogous. The out-of-time ordered correlators (OTOC),  which quantify the degree of non commutativity in
time between two operators, have been widely used as a common tool to investigate chaos in quantum systems \cite{Maldacena2016,Wang,Roberts2015,FAN2017707,Luitz,Herrera,Izrailev,Yan_2019}. In general the exponential growth rate of the OTOC is related to positive Lyapunov exponents \cite{Chavez2019}, which is interpreted as high sensitivity to initial conditions i.e. connected to the notion of chaos. However, the OTOC can also grow exponentially fast  in one-degree-of-freedom quantum systems that are not globally chaotic, but are critical \cite{Hummel,Chavez2019}. This is the case of the quartic double well potential which exhibits periodical trajectories but it contains a critical energy at the local maximum. 

In this manuscript we test different approximation methods in quantum mechanics in order to chose the most suitable option to access to highly excited states and compare them to a semiclassical method (EBK). Then we use the complete spectrum below the critical energy to observe classical correspondence in the density of states when $\hbar\to0$. Finally we  identify signatures of the  positive Lyapunov exponent on the spectrum.

\section{Generalities}

\begin{figure}
\centerline{\includegraphics[width=0.6\textwidth
]{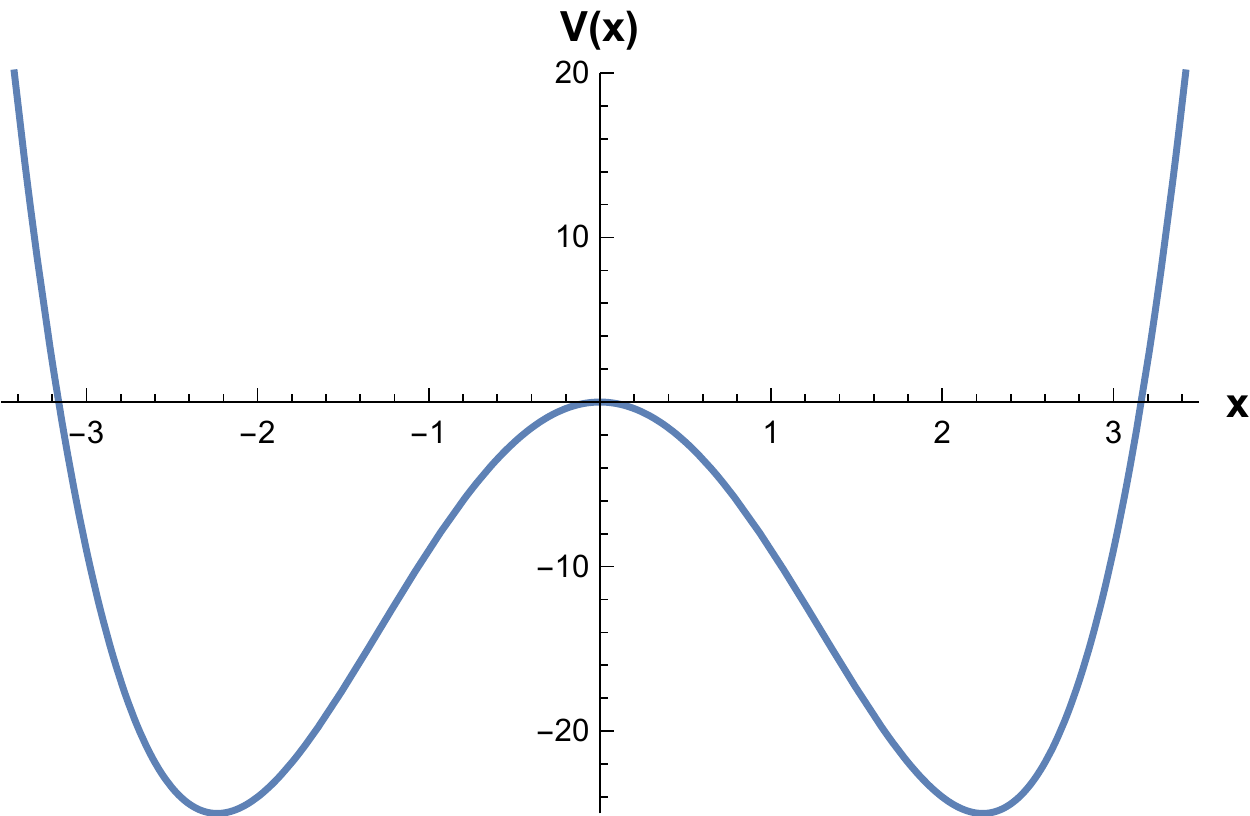}}
\caption{\label{pot} Double well potential in (\ref{Ham}) with constants  $a=-10$, $b=1$.}
\end{figure}

The Hamiltonian we are considering is the following
\begin{equation}
\label{Ham}
H=\frac{p^2}{2m}+a x^2+b x^4\,,
\end{equation}
where $p$ represents the momentum, $x$ the coordinate and $m$ the mass which, for simplicity, will be set from now on to the unity, $m=1$. In the quantum case, the variable $x$ and $p$ are promoted to operators $x\to\hat{x}$ and $p\to \hat{p}=-i\hbar\frac{d}{dx}\,$.
The potential is plotted in Fig. \ref{pot} for particular constants $a=-10$ and $b=1$. It shows degenerated minima at $x=\pm\sqrt{\frac{-a}{2b}}=\pm\sqrt{5}$ and a local maximum at the center of coordinates $x=0$ which represents an unstable point of equilibrium at a critical energy $E_c=0$. 

From a classical point of view, we can obtain the period of the orbits by   considering  action-angle coordinates
$$
J=\oint p dx= 2\int_{x_1}^{x_2} p dx= 2\int_{x_1}^{x_2} \sqrt{2 } \sqrt{E-V(x)}dx,
$$
where $x_i$ ($i=1,2$) are the classical turning points. The period in which the coordinates undergo an oscillation 
 is given by
$$\nu=\frac{1}{T}=\frac{\partial E}{\partial J},$$ 
and therefore the period for a particular value of the energy $E$ can be obtained by the following integral
\begin{equation}
\label{Tclasico}
T=\frac{\partial J}{\partial E}=\int_{x_1}^{x_2} \frac{\sqrt{2}}{\sqrt{E-V(x)}}dx\,.
\end{equation}

\begin{figure}
\centerline{\includegraphics[width=0.7\textwidth
]{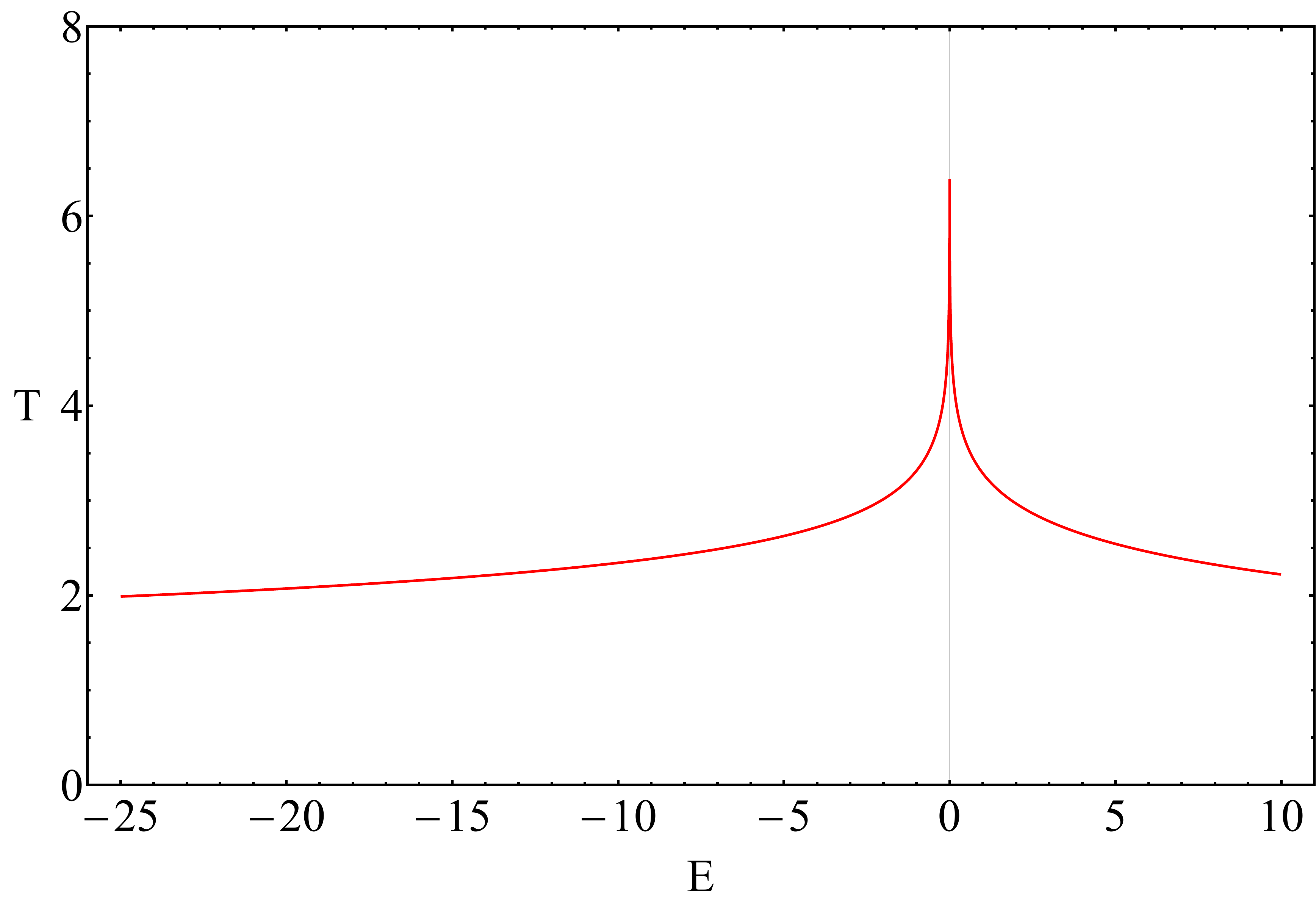}}
\caption{\label{periodo} Classical period $T$ as a function of the energy $E$. In order to obtain a symmetric curve around the critical energy $E_c=0$, the period for trajectories with  $E<0$ was multiplied by a factor~$2$. }
\end{figure}

In Fig. \ref{periodo} the classical period $T$ is plotted as a function of the energy. The divergence corresponds to the critical energy $E_c=0$, associated to the unstable equilibrium point at $x=0$, which physically implies that the particles at the double well potential take infinite time to reach the unstable equilibrium point.

In quantum mechanics, since the potential grows to infinite for $x\rightarrow \pm\infty$, there are infinite bound states, however the number of states with energy below the critical $E_c=0$ is finite. The Hamiltonian commutes with the parity operator, $\hat{P}\psi(x)=\psi(-x)$, consequently the energy eigenfunctions are symmetric ($P=1$) or anti-symmetric ($P=-1$ ), $\hat{P}\psi_k^\pm(x)=\pm\psi_k^{\pm}(x)$.
Below the critical energy,  the eigenstates organize in pairs ($P=\pm$)  of quasidegenerate levels.  The probability of tunneling between the two wells diminish as a function of the  energy gap between these quasi-degenerate pairs \cite{Bhatta1996,Yuste1993}, and is it weaker for states in the bottom of the wells.
The energy gap between parity partners  increases as their energy approaches the top of the double well, and so does   the probability of tunneling. The energy gap between the lowest levels has been studied to a certain depth \cite{Zhou,Park} for different separations between the two potential minima. This type of solution to the Schr\"odinger equation leads to the concept of instantons \cite{SHURYAK1988621,doi:10.1063/1.526205,Van}.

Within the approach of quantum mechanics, the  frequency and therefore the period of the quantum dynamics inside either well  at a certain  energy region is  given by  the energy difference of consecutive levels with the same parity  as
\begin{equation}
    E_{k+1}^{\pm}-E_{k}^{\pm} = \hbar \omega = \frac{2 \pi \hbar}{T}\,.
\end{equation}
By defining the energy density of states of a given parity  as $\rho^{\pm}(\bar{E}) = 1/(E_{k+1}^\pm-E_{k}^{\pm})$, where $\bar{E}=(E_{k+1}^\pm+E_{k}^{\pm})/2$, the period as a function of the energy is given by
\begin{equation}
    \label{Tcuantico}
    T= 2\pi \hbar \rho^\pm(\bar{E})\,.
\end{equation}
The total density of states, taking into account the quasi double degeneracy of energy-level pairs with different parity is $\rho(\bar{E})=2\rho^\pm(\bar{E})$, therefore 
\begin{equation}
\label{densityQ}
2 T= 2\pi \hbar \rho(\bar{E}),\ \ \ \ \ \ \  \text{(for $E<E_c=0$).}
\end{equation}
For states with energy above the double well,  $E>E_c=0$, the double  quasi-degeneracy is broken and the period and density of states are related as follows
\begin{equation}
  T= 2\pi \hbar \rho(\bar{E}),\ \ \ \ \ \ \  \text{(for $E>E_c=0$),}
\end{equation}
 where $\rho(\bar{E})$ is the total density of states given by 
$\rho(\bar{E}) = 1/(E_{k+1}-E_{k})$ with $E_k$ the energy of levels irrespective of their  parity.

Following the Bohr's correspondence principle, the results from quantum mechanics, as $\hbar\to 0$, should reproduce those from classical mechanics where discrete energy levels or tunneling do not exist. However as $\hbar$ decreases, the number of bound states below the critical energy increases as $1/\hbar$, even though the potential remains fixed. In this case, large matrices are needed in order to obtain the spectrum below the critical energy. Accelerating the convergence of the energy levels is a priority to access to highly excited states close to the critical energy.  
 In the following section we briefly describe approximation methods we used to obtain the spectrum from the ground state  up to  the critical energy $E_c=0$.

\section{Methods}

Within the framework of variational methods, the simplest choice 
for building the trial function is to consider a linear superposition in a certain complete and orthogonal basis 
$$\Psi(x)=\sum_{k=1}^Nc_k\psi_k(x)\,,$$
where $N$ is the size of the basis set and $\psi_i$ the basis functions. 
In the following subsections we briefly describe 
different basis.

\subsection{Sinc Method}

Within the Sinc method, the wavefunction is a linear superposition of sinc functions
\begin{equation}
\label{sinc}
\psi_k(\Omega,x)=\frac{\sin(\pi(x-k\Omega)/\Omega)}{\pi(x-k\Omega)/\Omega},
\end{equation}
where $\Omega$ is the spacing between maximas of neighbouring sinc functions and $k$ is an
integer index which controls the location of such maximas and takes values from $-k_{max}$ to $k_{max}$.

The matrix elements of the Hamiltonian evaluated in the set of sinc functions are
\begin{eqnarray}
\label{matrixSinc}
V_{kk^\prime}&=&V(k\Omega)\delta_{kk^\prime} \nonumber \\
T_{kk}&=& \frac{\pi^2}{6\Omega^2} \nonumber \\
T_{k\neq k^\prime}&=& \frac{(-1)^{k^\prime-k}}{\Omega^2(k^\prime-k)^2}\nonumber \\
\end{eqnarray}

The spacing parameter $\Omega$ is treated as variational and its optimal value is obtained by solving 
$\frac{d}{d\Omega} {\rm Tr}[\hat{H}]=0$. The size of the matrix representations is $N\times N$ where
$N=2k_{max}+1\,$ is the number of sinc functions. For a detailed description we refer to \cite{Amore_2006}.

\subsection{Lagrange Mesh Method}

In the LMM the wave function is composed by a superposition of Lagrange functions which satisfy the Lagrange condition
$\psi_k(x_{k^\prime})=\lambda_k^{-1/2} \delta_{kk^\prime}\,,$ thus facilitating the integration of the matrix elements using the Gauss quadrature.
In particular for the Hermite mesh, the Lagrange functions are 
$$\psi_k(x)=\frac{(-1)^{N-k}}{(2h_N)^{1/2}}\frac{H_N(x)}{(x-x_k)}e^{-x^2/2}\,,$$
where $H_N$ are the Hermite polynomials, $x_i$ the zeroes of $H_N$, $N$ the size of the mesh and $h_N=2^NN!\sqrt{\pi}$.
 The matrix elements for the potential and the kinetic terms are
\begin{eqnarray}
V_{kk^\prime}&=&V(x_i)\delta_{kk^\prime}\,,\nonumber\\
T_{kk}&=&\frac{1}{3}(2N+1-x_k^2)\,, \nonumber \\
T_{k\neq k^\prime}&=&(-1)^{k-k^\prime}\frac{2}{(x_k-x_k^\prime)^2}\,.
\end{eqnarray}
For details we refer to \cite{Amore_2006}. For numerical calculations one can use the Mathematica package~\cite{delValleLMM}.

\subsection{Hermite basis}

In this method the wave function is considered as a linear superposition 
composed by eigenfunctions of the harmonic oscillator i.e. Hermite polynomials weighted by a Gaussian function \cite{Verguilla}.
The simplest way to construct the matrix representation in this basis, is rewriting the Hamiltonian 
(\ref{Ham}) in terms of the usual ladder operators $\hat{a}= \frac{1}{\sqrt{2\hbar}}(\hat{x}+i\hat{p})$ and $\hat{a}^\dagger=\frac{1}{\sqrt{2\hbar}}(\hat{x}-i\hat{p})$
acting on eigenstates of the  harmonic oscillator $|n\rangle$. However,  the convergence rate of the energy in the truncated basis can be accelerated by introducing a scaling $x^\prime\to \Omega x$ \cite{AMORE200587}, which in turns transform the annihilation  and  creation operators as $$\hat{a}_\Omega= \frac{1}{\sqrt{2\hbar}}\left(\frac{\hat{x}}{\Omega}+i\Omega\hat{p}\right)$$ and $$\hat{a}^\dagger_\Omega=\frac{1}{\sqrt{2\hbar}}\left(\frac{\hat{x}}{\Omega}-i\Omega\hat{p}\right),$$ respectively. The scaling parameter is optimized by $\frac{d}{d \Omega}{\rm Tr} \hat{H}=0$.
In terms of scaled operators, the Hamiltonian reads
$$\hat{H}=-\frac{\hbar}{4\Omega^2}(\hat{a}^\dagger_\Omega-\hat{a}_\Omega)^2+b\left(\frac{\hbar}{2\Omega^2}\right)^2(\hat{a}^\dagger_\Omega+\hat{a}_\Omega)^4+\frac{a\hbar}{2\Omega^2}(\hat{a}^\dagger_\Omega+\hat{a}_\Omega)^2\,.$$
Thus the matrix representation of the Hamiltonian is tridiagonal with non-zero elements
\begin{eqnarray}
\label{HNN}
H_{n,n}&=& \frac{b\Omega^4}{4}(6n^2 + 6n + 3)\hbar^2+\left(\frac{a\Omega^2}{2} + \frac{1}{4\Omega^2}\right)(2n + 1)\hbar\nonumber \\
H_{n-2,n}&=&\sqrt{n(n-1)}\left( \frac{b\Omega^4}{4} (4n - 2)\hbar^2 +\frac{a\Omega^2}{2 }\hbar -\frac{\hbar}{4\Omega^2}\right) \nonumber \\
H_{n-4,n}&=&\frac{b\Omega^4}{4 }\hbar^2\sqrt{n(n-1)(n-2)(n-3)}\,,\nonumber \\
\end{eqnarray}
and their corresponding  transposed elements. The basis is truncated at $n_{max}$ and the size of the matrix is $N\times N$ where $N=n_{max}+1$.

\subsection{Einstein-Brillouin-Keller}

This method, named after Einstein-Brillouin-Keller \cite{KELLER1958180} (EBK), belongs to the semiclassical methods since it is an improvement to the Bohr-Sommerfeld quantization. The quantization rule is the following

\begin{equation}
    J =  \oint p dx = 2\pi \hbar \left( n + \frac{\mu}{4} + \frac{d}{2} \right)\,,
\end{equation}
\\
where $J$ is the angle-action variable, $n$ is a non-negative integer (the quantum number), $\mu$ and $d$ are the Maslov index:
$\mu$ corresponds to the number of  turning points in the classical trajectory ($\mu=2$  in this case) and $d$ corresponds to the number of reflections with a hard wall (absent in this case $d=0$).
Since EBK is a semiclassical method, it is expected to correctly reproduce the spectrum for large quantum number $n$ \cite{Hruska} and therefore to be a good approximation to access to highly excited states when $\hbar\to 0$.

\section{\label{wf} Results}

We performed numerical diagonalization of the Hamiltonian (\ref{Ham}) for different values of $\hbar$ in order to approach the classical limit and identify some signatures of classical instability in the spectrum of the quartic oscillator. Our results are organized in four subsections.
For simplicity we  focus our attention to a 
quartic potential with constant values $a=-10$ and $b=1$.

\subsection{Convergence}

\begin{figure*}
    \begin{tabular}{cc}
    \hline \\
    $(\hbar=1,n=1)$ & $(\hbar=1,n=5)$\\
    \includegraphics[width=8cm, height=4cm,angle=0]{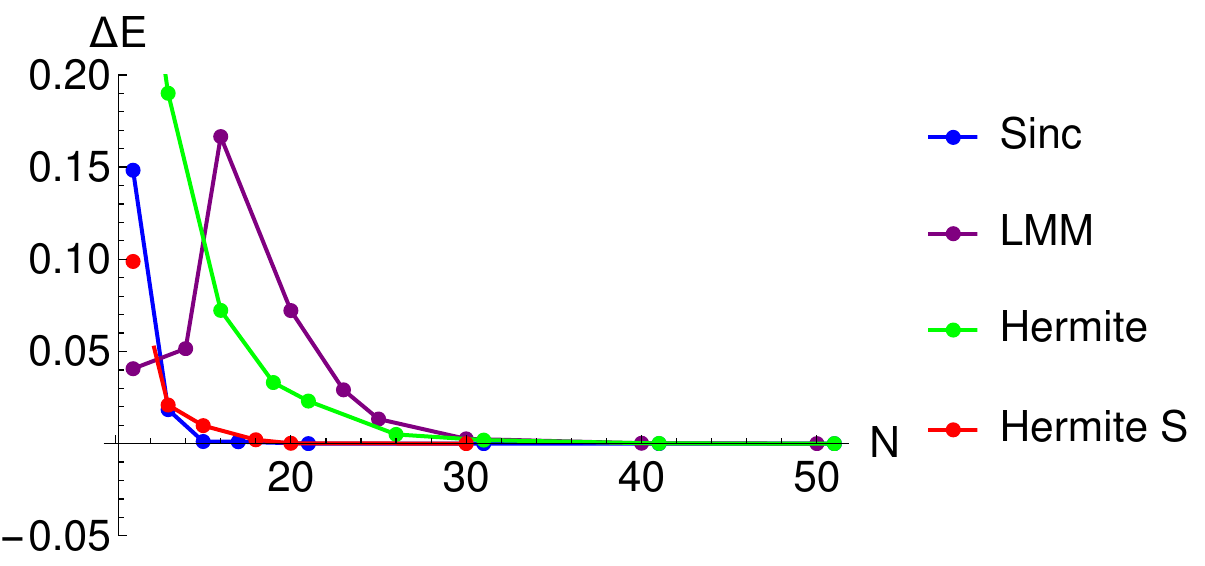}  &
     \includegraphics[width=8cm, height=4cm,angle=0]{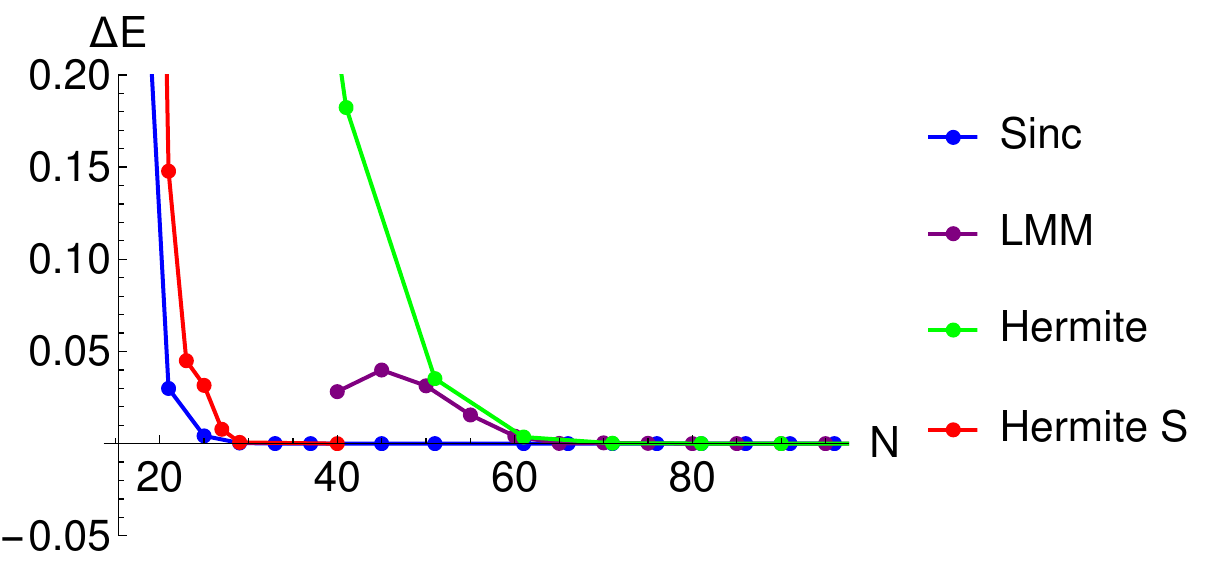}  \\
     $(\hbar=\frac{1}{10},n=20)$ & $(\hbar=\frac{1}{10},n=50)$\\
      \includegraphics[width=8cm, height=4cm,angle=0]{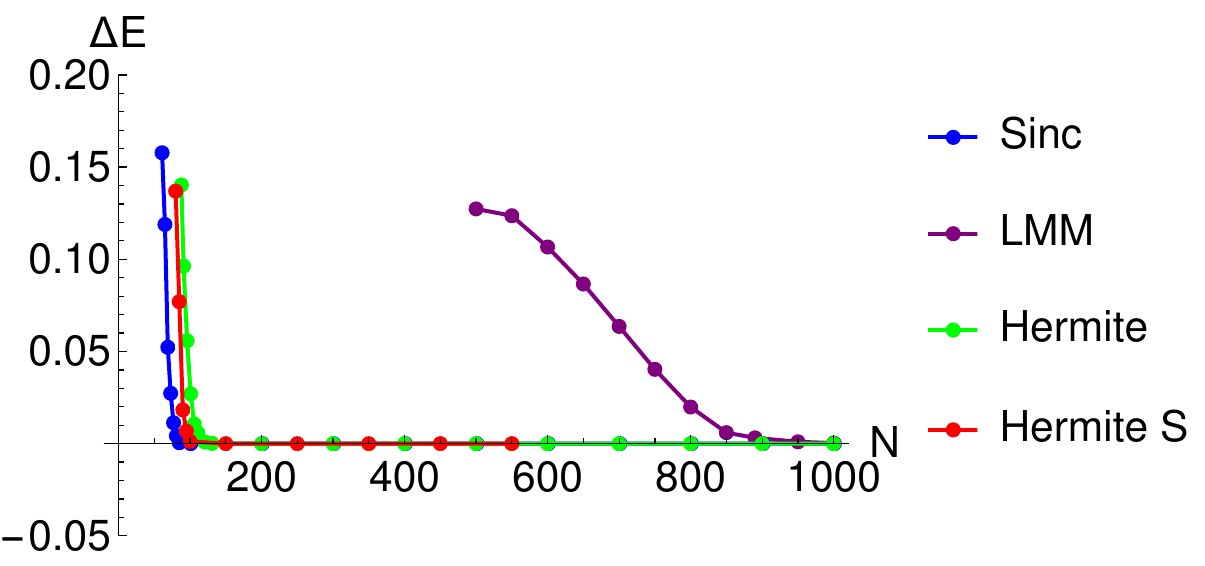} &
          \includegraphics[width=8cm, height=4cm,angle=0]{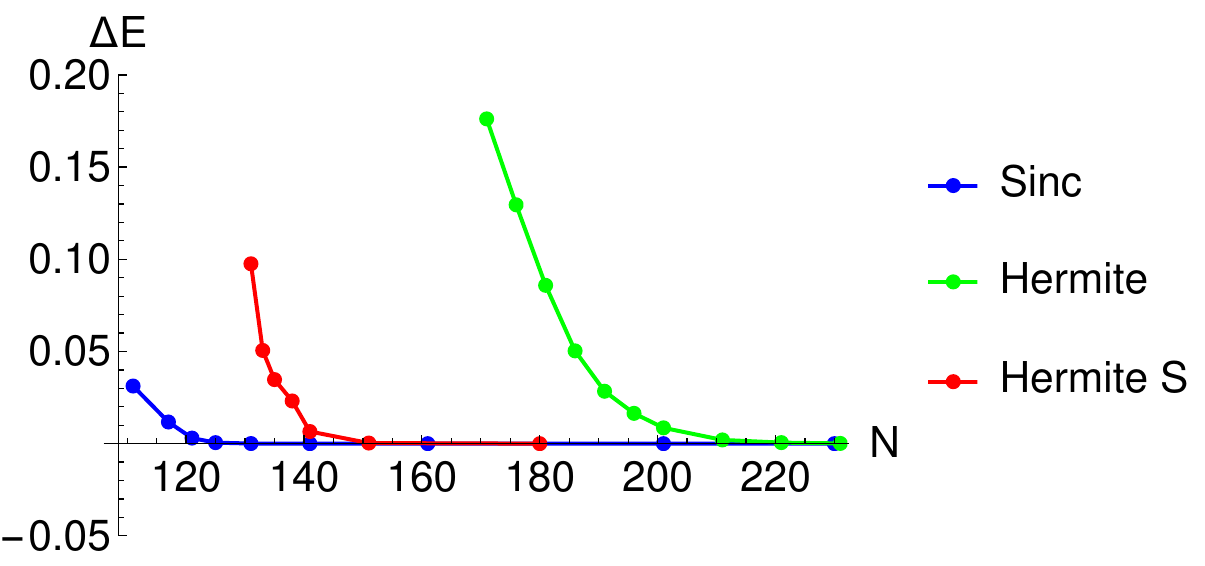}  \\
    $(\hbar=\frac{1}{50},n=100)$ & $(\hbar=\frac{1}{100},n=200)$\\
      \includegraphics[width=8cm, height=4cm,angle=0]{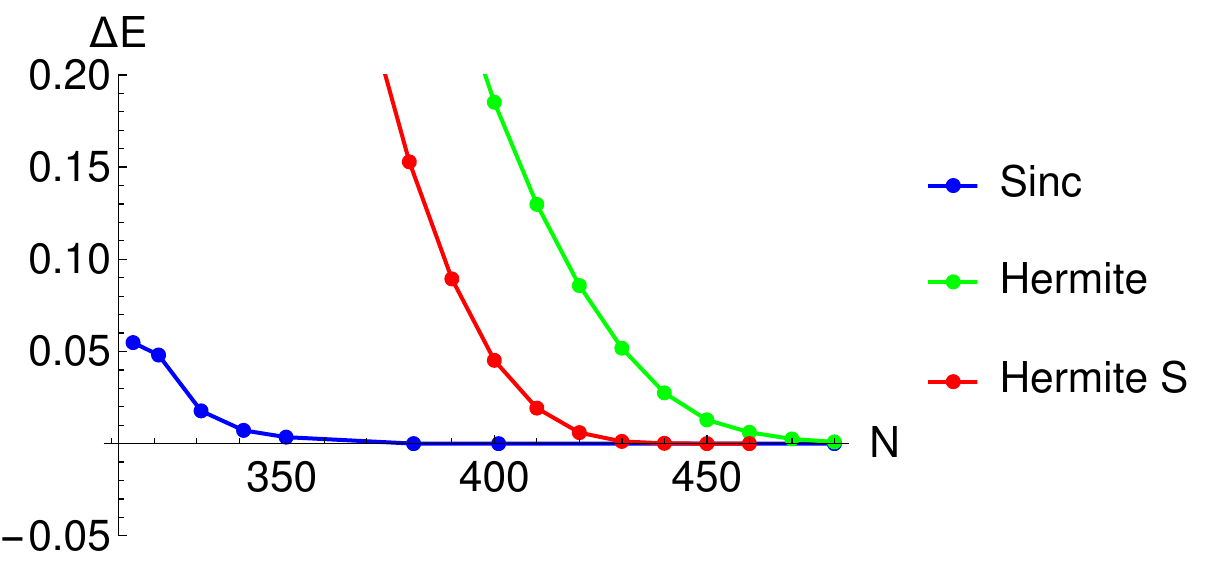} &
          \includegraphics[width=8cm, height=4cm,angle=0]{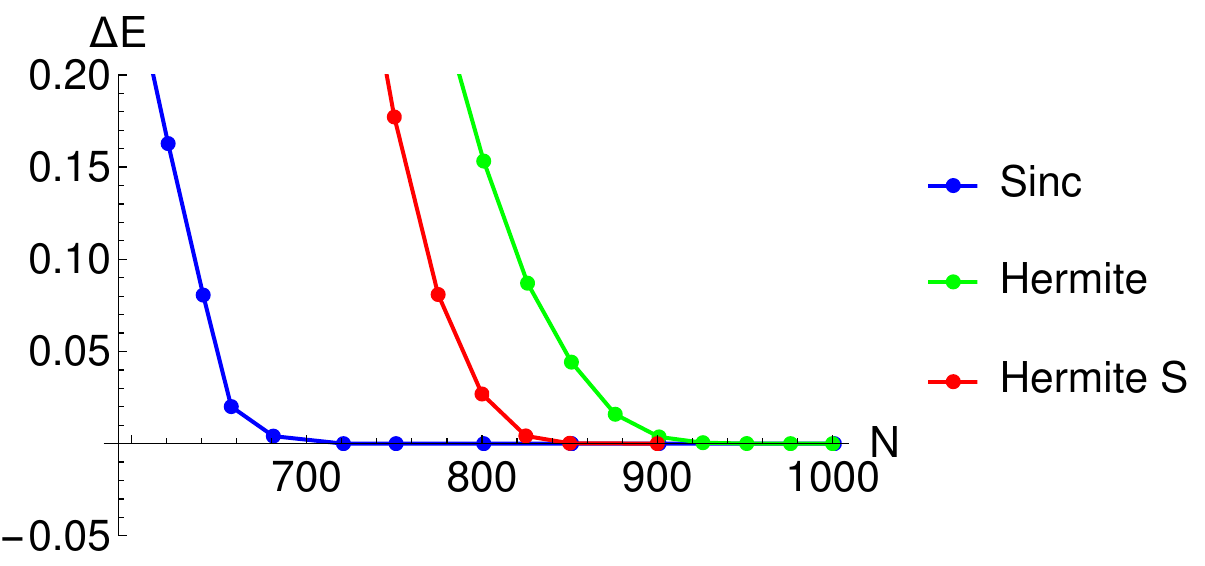}  \\
\hline
    \end{tabular}
    \caption{\label{convergence}
    Convergence of the energy $\Delta E$ as a function of the size of the basis $N$ for the following basis: Sinc (blue), LMM (yellow), Hermite (green) and Hermite with scaled coordinate (red).
    }
\end{figure*}

We tested the convergence of the energy for different levels and different values of $\hbar$ as a function of the basis size $N$ for the following basis: (i) Sinc, (ii) LMM basis, (iii) Hermite functions and (iv) the Hermite functions with scaled coordinate. We measure the rate of convergence by computing the energy difference $\Delta E_n{(N)}=|E_n(N)-E^{ref}_n|$ where $n$ is the quantum number, $N$ the size of the basis and $E^{ref}_n$ a reference energy. We use as reference the results obtained by the corresponding method with the basis size $N=2000$.
Since the three methods involve numerical diagonalization, we use the routine    \textit{diasym} of LAPACK and perform diagonalization in \textit{fortran90}. Obtaining the matrix representation does not imply a major difference in computing time for any method. As for the optimization of the scaling parameters $\frac{d}{d \Omega}Tr\hat{H}=0$ it is necessary to solve a third (Hermite) or  sixth-degree (Sinc) equation in $\Omega$. This process does not represent a significant addition on computational time even for large basis size.  However LMM requires to obtain $N$ roots of the Hermite polynomials to build the mesh thus increasing considerably the computing time for large values of $N$. 

\begin{figure}
 \centerline{\includegraphics[width=0.6\textwidth
 ]{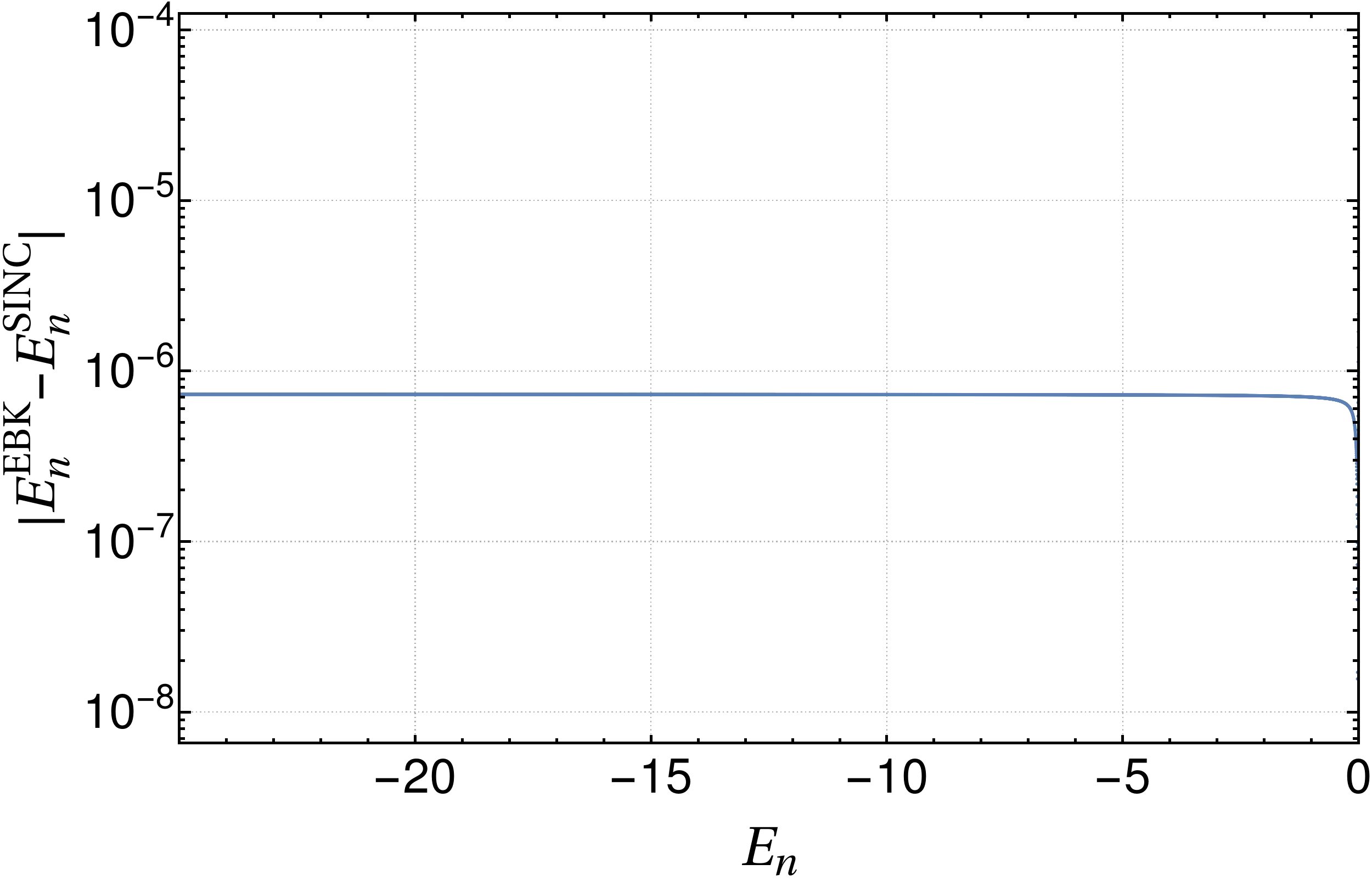}}
\caption{\label{SincvsEBK} Energy difference between the Sinc Method and EBK as a function of the energy of the state $n$ (obtained by using Sinc basis), corresponding to $\hbar=1/2000$.}
\end{figure}

In Fig. \ref{convergence} we plot $\Delta E_n(N)$ as a function of the basis size considering the four basis. One can appreciate that 
the fastest convergence is always obtained with the Sinc basis, while the  lowest is for LMM. One can notice that from $\hbar=\frac{1}{10}$ the convergence rate of Lagrange basis is very slow. This is mainly because the mesh points must be redistributed by introducing a scaling parameter (see \cite{BAYE20151}) and treating it as variational parameter. In LMM there is no  closed expression to estimate its optimal value, instead numerical minimization must be performed to obtain its optimal value. Since the optimal value of the scaling parameter  highly depends on the basis size $N$ and the quantum number $n$, the analysis is impractical for highly excited states. On the other hand, the optimization of the scaling parameter $\Omega$, corresponding to the spacing between maximas within the Sinc and squeezing within  the  Hermite basis, accelerates the rate of the convergence by using the closed expression $\frac{d}{d\Omega}{\rm Tr}{\hat{H}}=0$. 
Finally we compared the accuracy of the Sinc method to the semiclassical method EBK. It can be seen from Fig. \ref{SincvsEBK}
that EBK provides correctly at least six decimal digits in the entire region below the critical energy $E_c=0$ for $\hbar=\frac{1}{2000}$.

\subsection{Quantum-classical correspondence}

As $\hbar\to0$ it can be expected that the energy difference between discrete states vanishes. Therefore more bound states are contained below the critical energy even though the potential remains fixed.  The number of states which lie below the critical energy can be counted using the EBK results and they can also be corroborated with the Sinc method.  In Table \ref{tablahbar}, the number of bound states with energy below the critical for different values of $\hbar$ is presented.

Whereas  the EBK does not have a limitation for the number of bound states, the quantum methods are limited by the size of the matrix to perform diagonalization. 
The minimum value reached in this work is $\hbar=\frac{1}{2000}$ where  18980 bound states lie below the critical energy.

\begin{figure*}
    \begin{tabular}{ccc}
    \hline \\
    $\hbar=\frac{1}{10}$ & & $\hbar=\frac{1}{100}$  \\
    \includegraphics[width=0.45\textwidth]
    {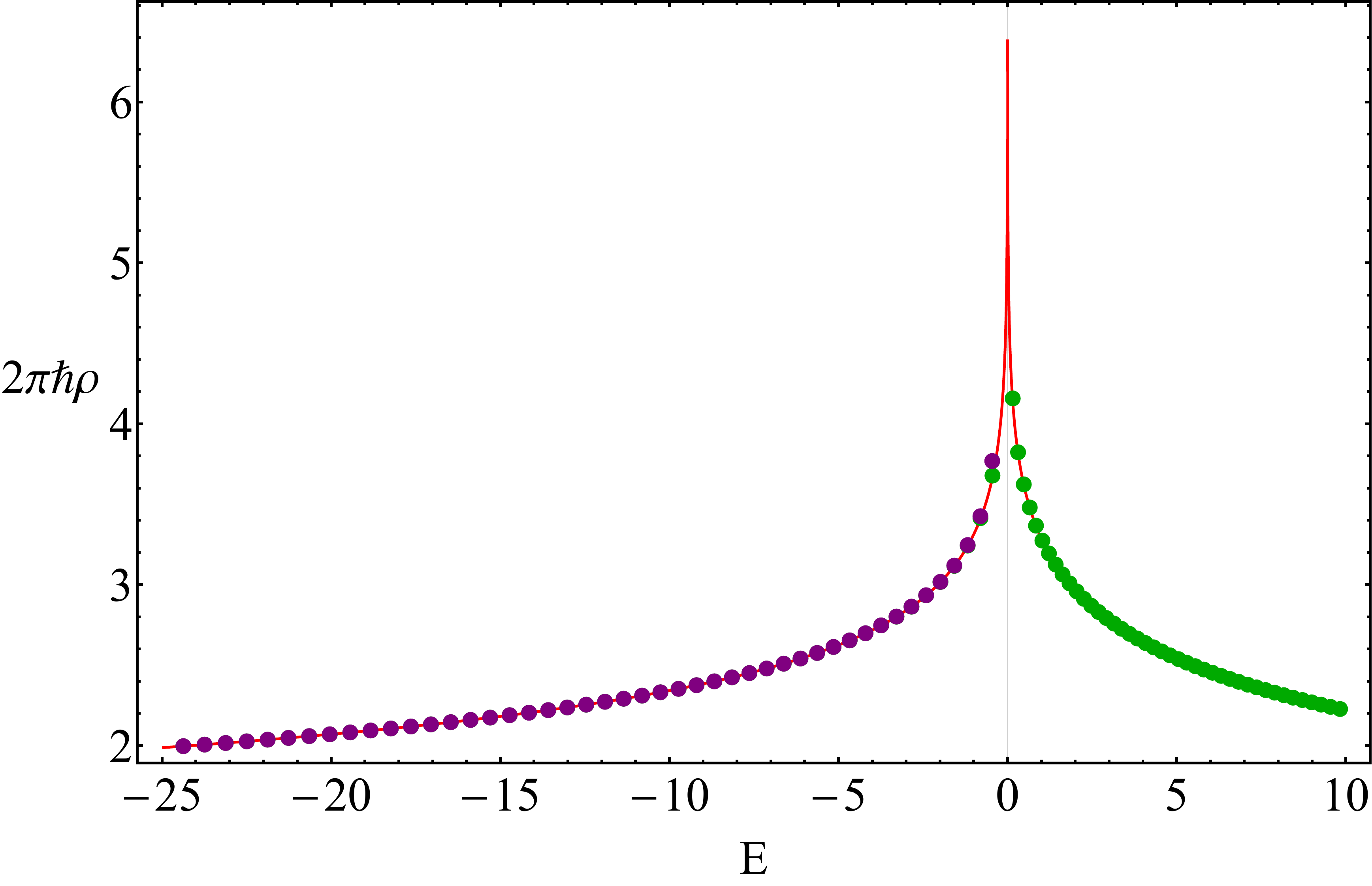} 
   & & \includegraphics[width=0.45\textwidth]
   {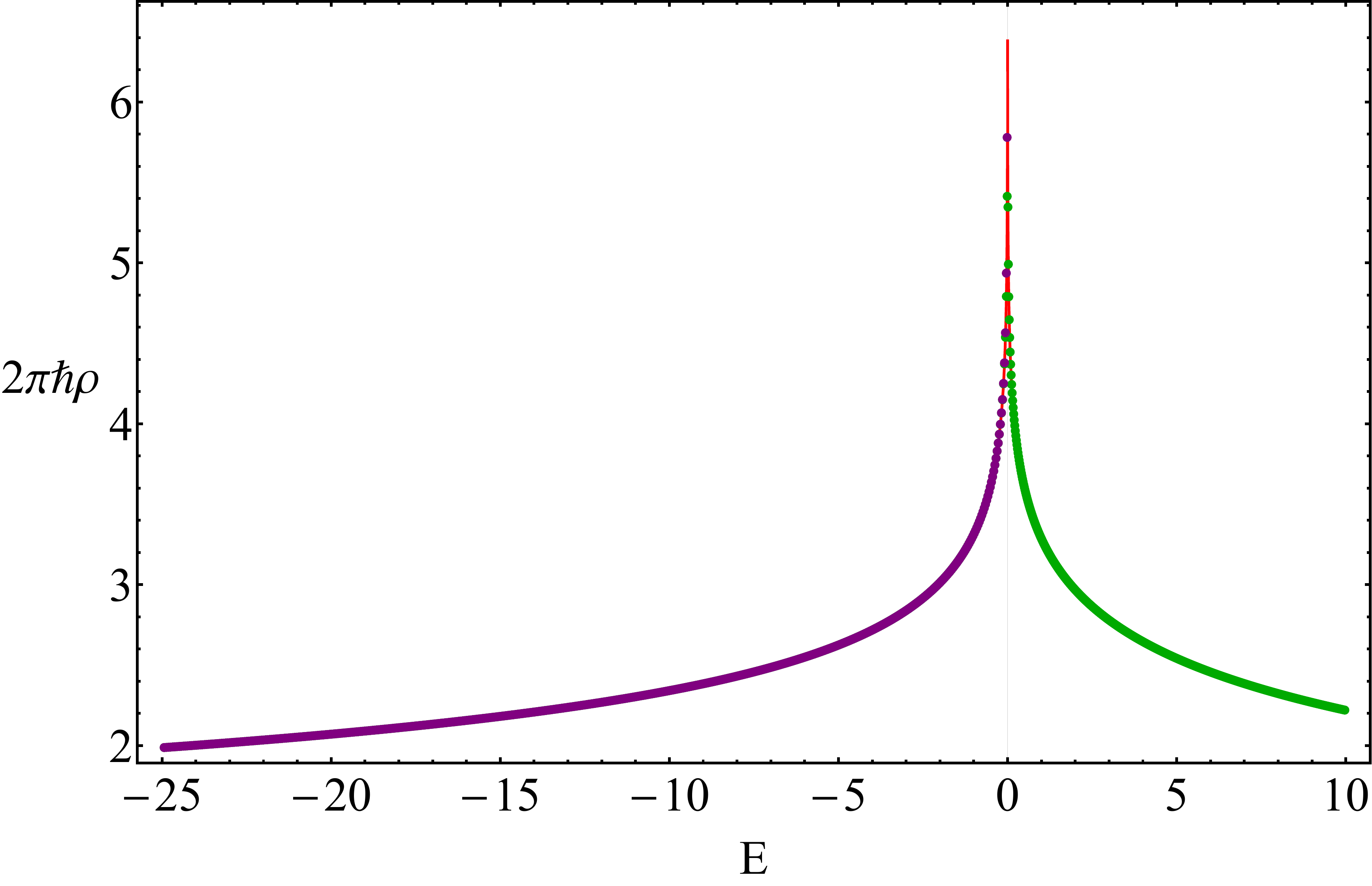} \\
    $\hbar=\frac{1}{1000}$ & & $\hbar=\frac{1}{2000}$  \\
  \includegraphics[width=0.45\textwidth]
  {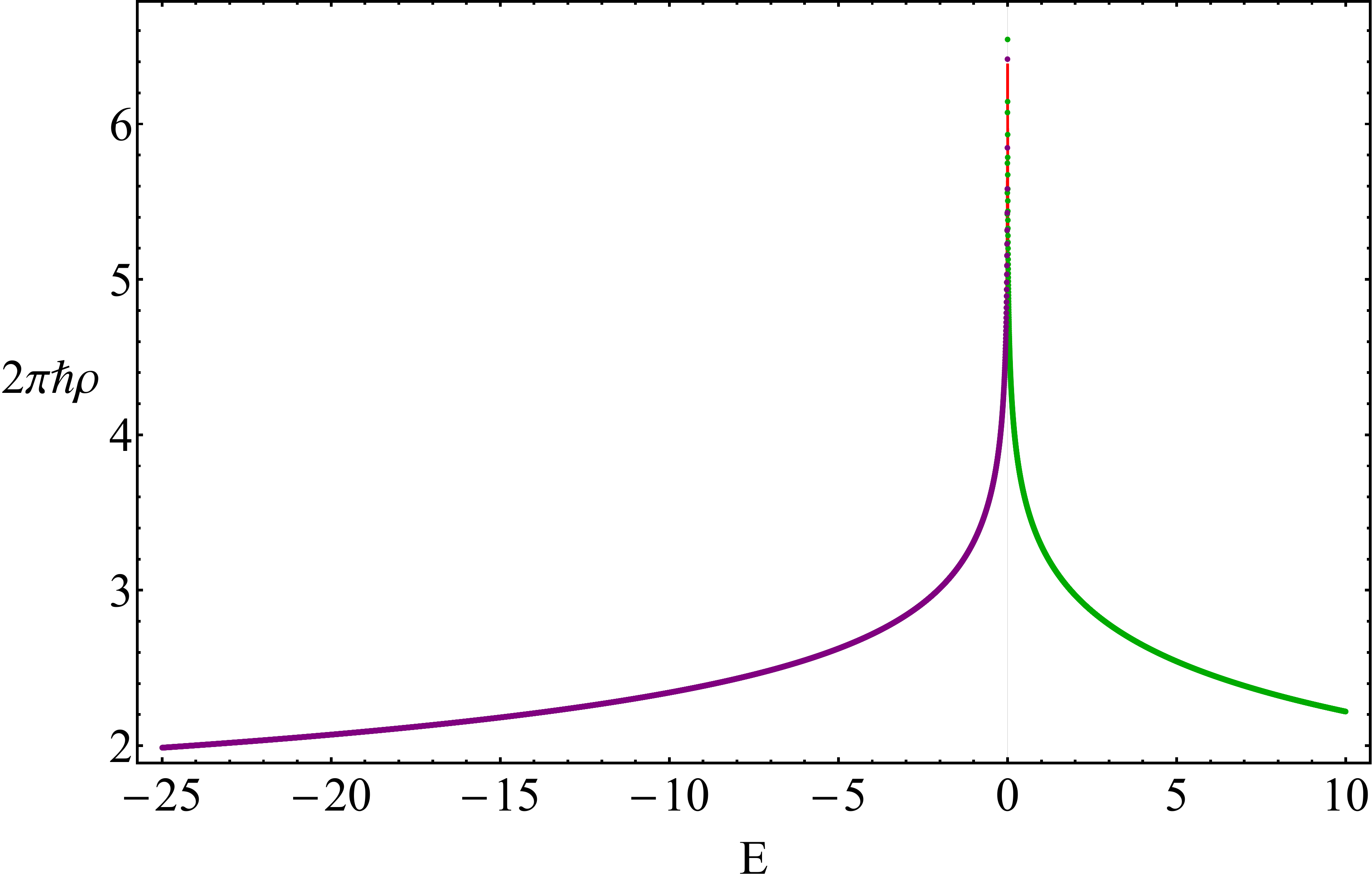} 
   &  &\includegraphics[width=0.45\textwidth]
   {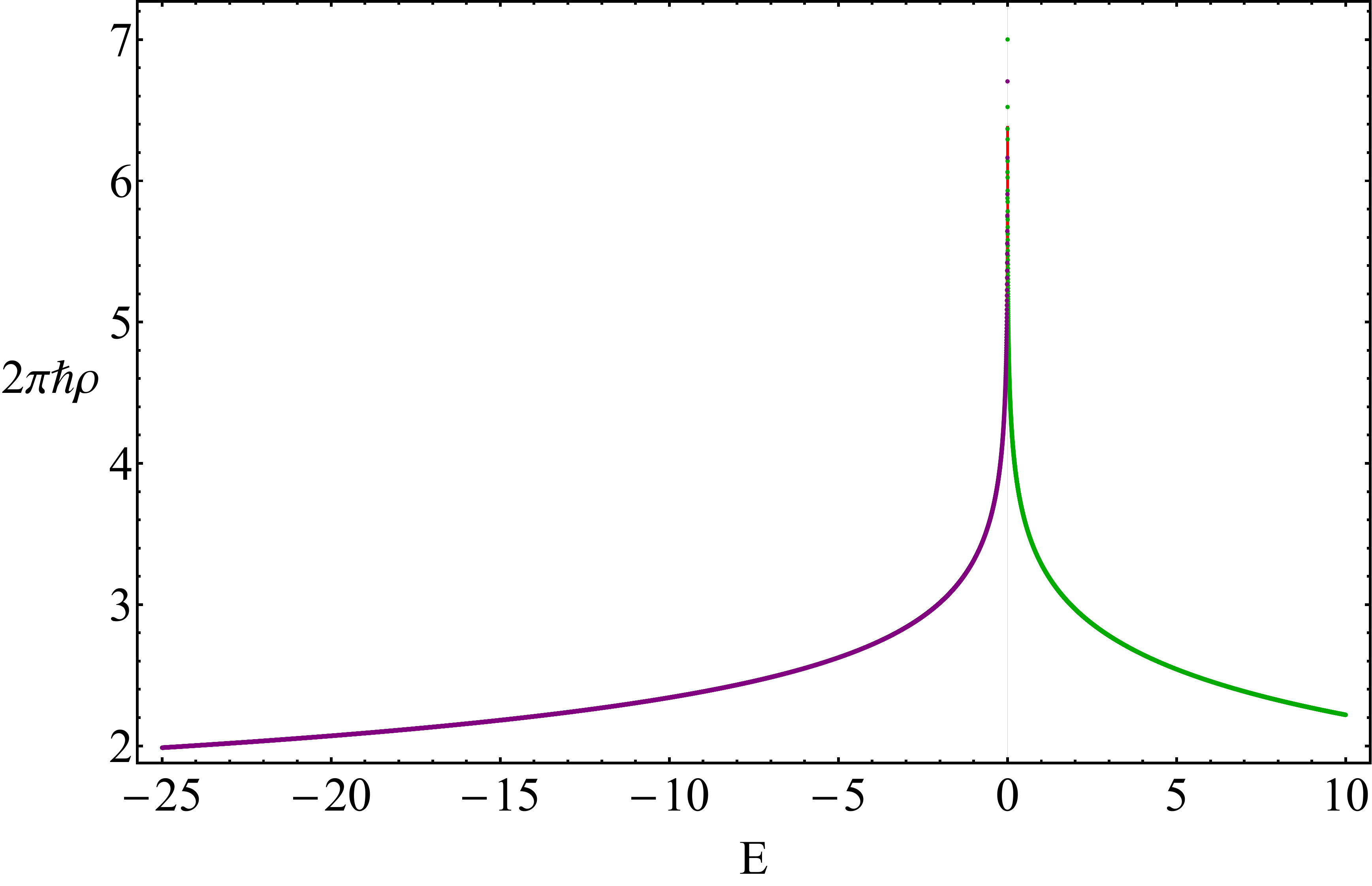} \\
\hline
    \end{tabular}
    \caption{\label{Correspondence}
    Dots depict the energy density of states, $2\pi\hbar\rho$,  as a function of the energy for different values of $\hbar$. Solid red line is the classical period for $E>0$ and twice the period for $E<0$. Purple dots indicates quantum density of states using energy levels from Sinc method and green dots using levels from EBK. 
    Before the critical energy, green dots lie behind the purple dots and only a few of them can be appreciated.
    }
\end{figure*}

In Fig. \ref{Correspondence} we plot the continuous classical period Eq.(\ref{Tclasico}) along with the discrete quantum density  Eq.(\ref{densityQ}),  for different values of $\hbar$. We notice that the discrete points from the quantum mechanics are distributed along the classical continuous period and the emergence of the divergence corresponding to the critical energy becomes evident as $\hbar\to 0$. Before the critical energy $E<E_c=0$,  according to Figure \ref{SincvsEBK}, the difference between the energies obtained by diagonalization in the Sinc basis and the EBK is of the order $10^{-6}$ energy units, because of that in Fig.\ref{Correspondence} the EBK values (green dots) lie behind the results coming from the Sinc method(purple dots) and only a few of them can be appreciated in the region $E<0$.

\begin{widetext}
 \begin{center}
\begin{table}
\begin{center}
\begin{tabular}{c|ccccccccccccc} 
\hline
$\hbar$ & 1 & & $\frac{1}{10}$ & & $\frac{1}{100}$ & &$\frac{1}{200}$ & & $\frac{1}{500}$ & & $\frac{1}{1000}$ & &$\frac{1}{2000}$ \\ 
No. states  & 10 & & 94 & & 950 & & 1898 & & 4746 & & 9490 & & 18980 \\
 \hline \end{tabular}
\caption{\label{tablahbar} Number of states below the critical energy $E_c$ for different values of $\hbar$. 
}
\end{center}
\end{table}
\end{center}
\end{widetext}

\subsection{Tunneling decay}

\begin{figure*}
    \begin{tabular}{cc}
    \hline \\
    (a)& (b)\\
    \includegraphics[width=8cm, height=5cm,angle=0]{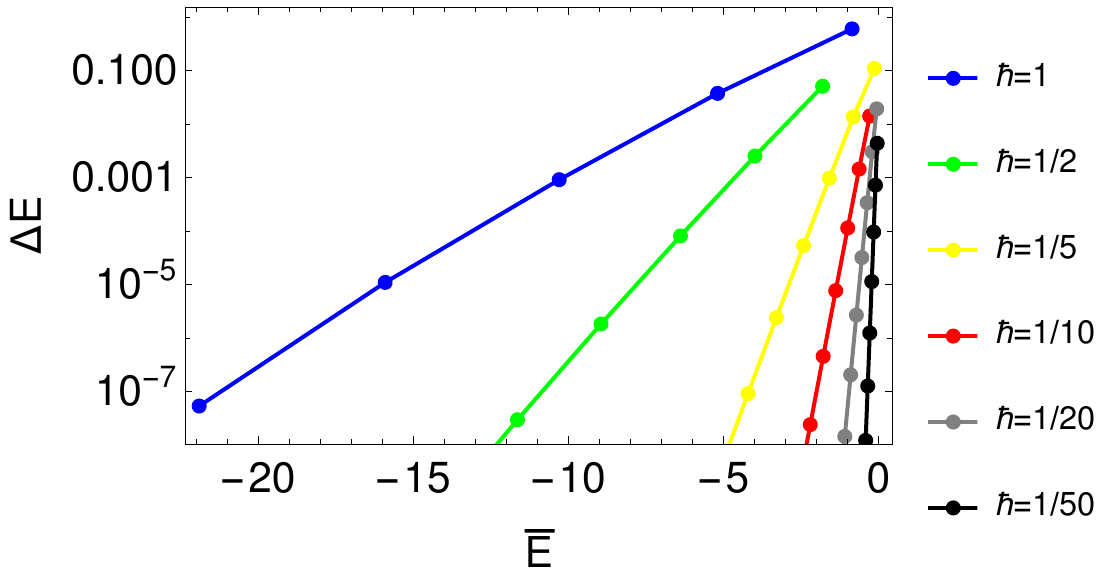}  &
     \includegraphics[width=8cm, height=5cm,angle=0]{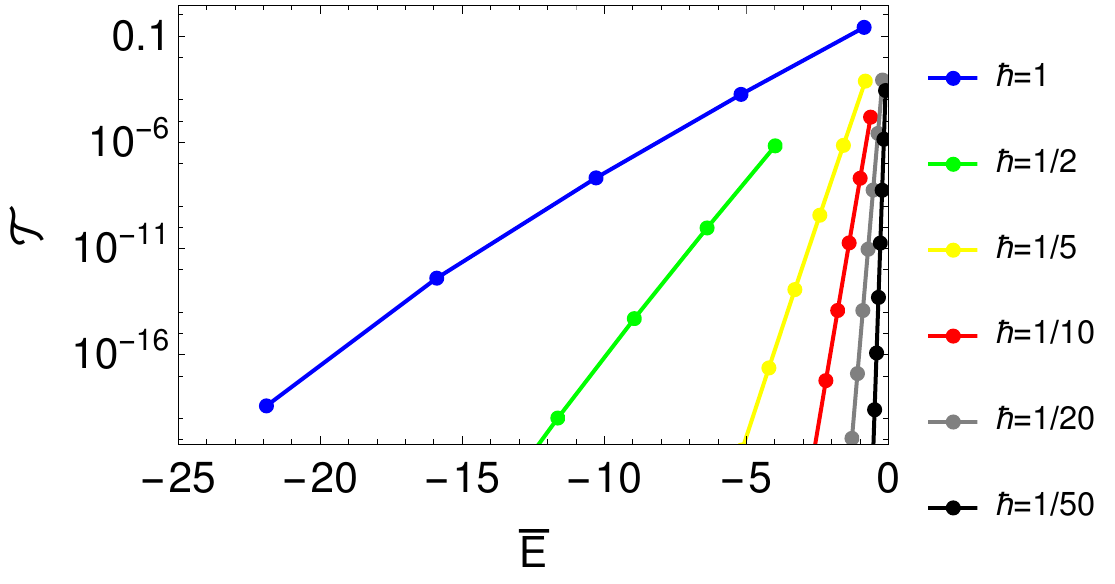}  \\
     \multicolumn{2}{c}{(c)}\\
   \multicolumn{2}{c}{\includegraphics[width=8cm, height=5cm,angle=0]{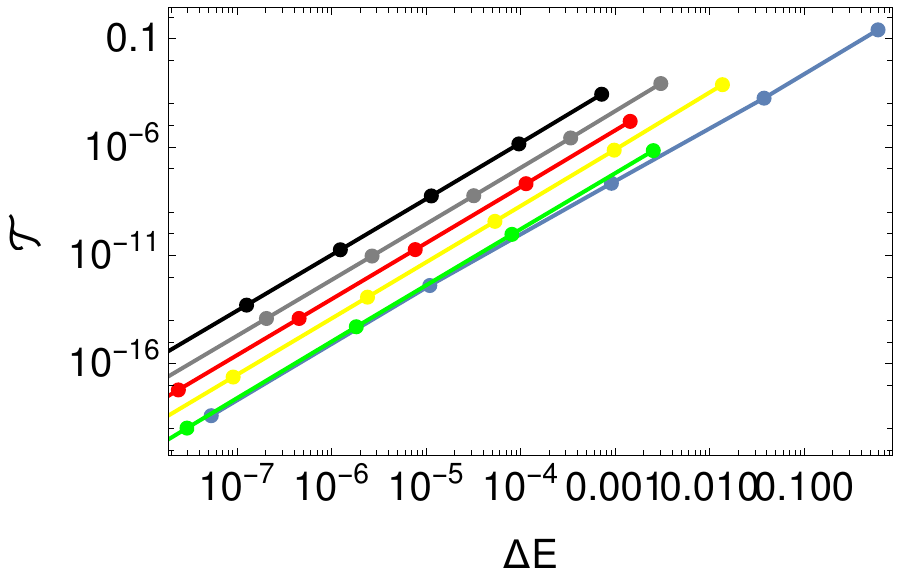}} \\
\hline
    \end{tabular}
    \caption{\label{GapT}
    (a)Energy gap $\Delta E$ between pairs of quasi-degenerate  energy levels of different parity and (b) coefficient of transmission $\mathcal{T}$ as a function of the energy average $\bar{E}$ for different values of $\hbar$. In  panel (c), the relation between $\mathcal{T}$ and $\Delta E$ is shown in $\log-\log$ scale. The same slope for the different lines implies that these quantities are related as $\mathcal{T}\propto (\Delta E)^\alpha$, while the different intercepts lead to $\mathcal{T}\propto \hbar^{-\alpha}$, where $\alpha=5/2$. 
    }
\end{figure*}

One of the most intriguing properties of the quantum theory is the possible quantum tunneling in the region of the double well which is classically forbidden. The energy gap between quasi-degenerated pairs   of different parity is a consequence of the tunneling phenomenon which appears in the energy region inside the double well. It is well known that the phenomenon of tunneling is less likely to be observed if the wells are too separated from each other \cite{Zhou}. 
In this case we kept the potential fixed but the tunneling effect in quantum mechanics should vanish when approaching the classical limit as $\hbar\to0$. Our high accurate results, provided by the Sinc method, give us access to the energy gap between two consecutive energy levels corresponding to different parities. On the other hand we can also use our high accurate energies to estimate the transmission coefficient by using the WKB approximation
$$\mathcal{T}=\exp\left( -2\sqrt{\frac{2}{\hbar}}\int_{x_1}^{x_2} \sqrt{|E-V(x^\prime)|}dx^\prime\right)\,$$
where $x_1$ and $x_2$ are the classical turning points.

In Figure \ref{GapT} we plot the energy gap of parity-partners energy levels $\Delta E$ at panel (a),  and the transmission coefficient $\mathcal{T}$ at panel (b) as a function of their energy average $\bar{E}$ for different values of $\hbar$. It can be noticed that the energy gap and the transmission coefficient decreases exponentially as the energy becomes more negative and the levels go deeper into  the double well.  This can be understood because the energy barrier is narrower close to the local maximum at $x=0$ corresponding to the critical energy $E=0$. As $\hbar$ decreases, the energy gap and transmission coefficient become detectable only for energies close to the  critical energy $E_c=0$.  Indicating, thus, that tunneling phenomenon can only be observed close to the critical energy when approaching  the classical limit. Given the linear behavior of the transmission coefficient and the energy gap of parity partner in the log-log scale, at Figure \ref{GapT} panel (c), we are able to establish that the tunneling increases as a function of the energy gap as ${\mathcal T}\propto(\Delta E)^\alpha$ and decay as ${\mathcal T}\propto\hbar^{-\alpha}$, where $\alpha=5/2$.


\subsection{Manifestation of the positive Lyapunov exponent}

\begin{figure}
\centerline{\includegraphics[width=0.6\textwidth
]{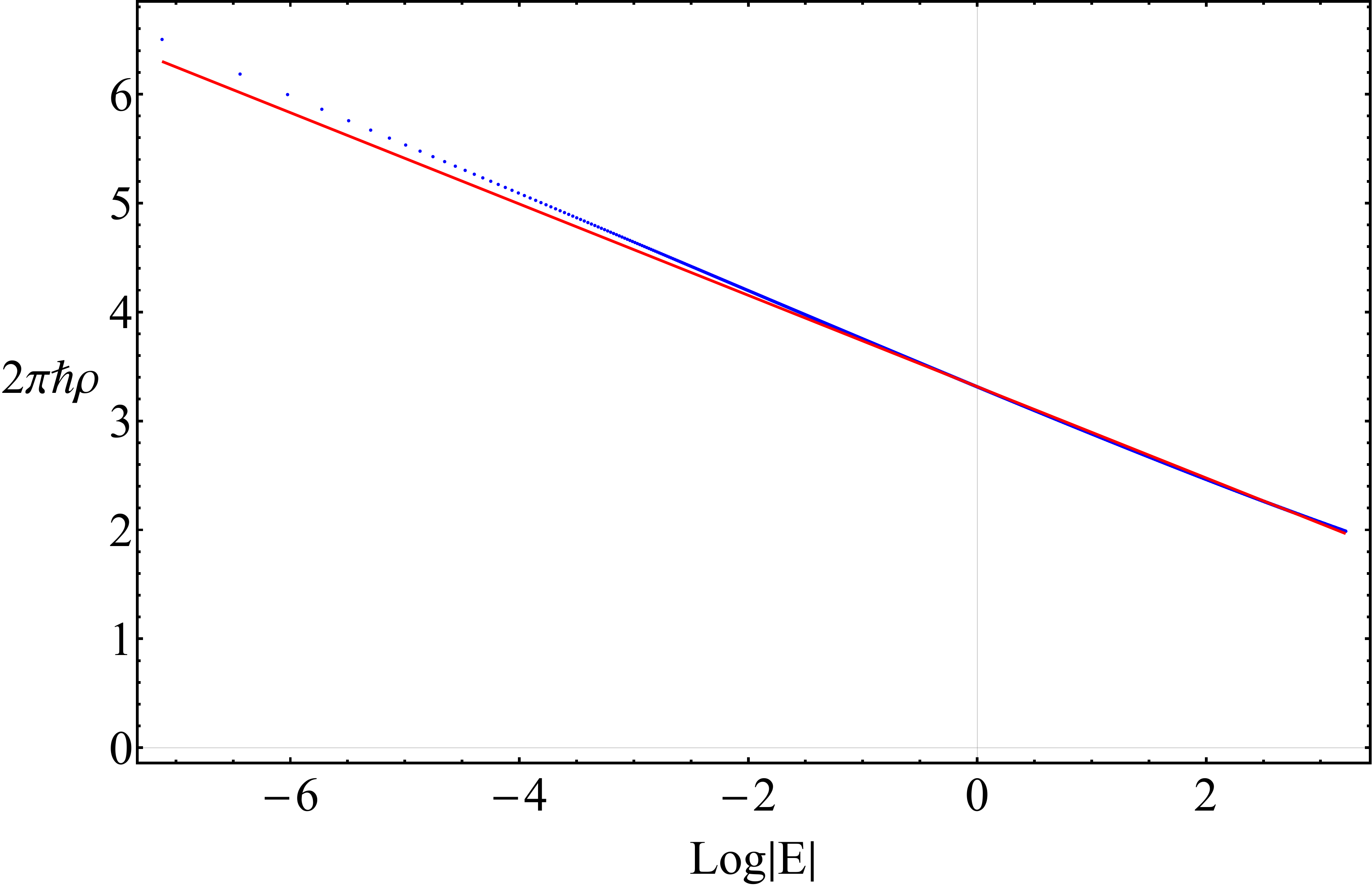}}
\caption{\label{PeriodvsLogE} Period $2\pi\hbar\rho(E)$ as a function of the logarithm of the energy $E$ with $\hbar=\frac{1}{2000}$. The red line correspond to the linear fit and the points to the numerical data.}
\end{figure}

From a classical treatment of the Hamiltonian (\ref{Ham}) it can be easily shown that the positive Lyapunov exponent which characterizes the critical behavior of the systems at the stationary point $(x_s,p_s)=(0,0)$ $(\dot{x}=0,\dot{p}=0)$ is (see Appendix~\ref{LyapunovAp} and \cite{PilatowskyChavez}) 
\begin{equation*}
\label{Lyapunov}
\lambda=\sqrt{-2 a}
=\sqrt{20}\approx 4.47214\,.
\end{equation*}


On the other hand, it can be shown that the density of states, close to the critical energy, behaves as 
\begin{eqnarray}
\label{linearrelation}
2\pi\hbar\rho(E)&=& 
-\frac{2}{\lambda} \log |E| - \frac{4}{\lambda} \log\left( \frac{\sqrt{b}}{-4 a} \right)\\
& =&-\frac{2}{\lambda} \log |E| - \frac{4}{\lambda} \log\left( \frac{1}{40} \right)\,,\nonumber 
\end{eqnarray}
where $\lambda$ is the Lyapunov exponent. In Appendix \ref{FormulaLogaritmica} we discuss the derivation of (\ref{linearrelation}) in detail.
Thus, the relation between the density of states and the logarithm of the energy is linear and the Lyapunov exponent $\lambda$ governs the slope  in agreement with the results from  Ref.\cite{Richter2019}.

In Fig. \ref{PeriodvsLogE} the density of states, for energies lower but very close to $E_c=0$, is plotted against the logarithm of the energy for the smallest value of $\hbar=1/2000$. We can observe that the data fits to a linear function with negative slope $\beta=-0.4192$. The slope given by the fit is similar to that in Eq. (\ref{linearrelation}) $\beta=-\frac{2}{\lambda}= -0.447214$ in terms of the Lyapunov exponent. This is an indication that the positive Lyapunov exponent governs the behavior of the quantum density of states close to the critical energy.  We think that the deviation of the slope would vanishes for smaller values of the Planck constant $\hbar$ than those reached in this work.

\section{Conclusions}

The Sinc method accelerate the rate of convergence of the energy levels
, allowing us to explore highly excited states close to the critical energy when approaching the classical limit. The optimization of spacing parameter $\Omega$ plays an essential role to increase the rate of convergence. The highly accurate results by the Sinc method allow to explore how the tunneling phenomenon decay when approaching the classical limit and how it can be detected only close to the critical energy.
On the other hand, the semiclasical method EBK provides at least six correct decimal digits for the entire energy region of interest in this work, thus facilitating the analysis of the density of states.

We observed that the highly excited states converge to the critical energy forming a logarithmic divergence in the density of states
. The quantum results go to the classical results as $\hbar\to0$ and they successfully reproduce the divergence on the classical period corresponding to the stationary point for the minimum value considered in this work $\hbar=1/2000$.
Furthermore, we found a manifestation of the positive Lyapunov exponent on the density of states of the spectrum of the quartic harmonic oscillator. Therefore we can conclude that the Lyapunov exponent  governs the logarithm divergent  behavior of the density of states for energies close to the critical $E_c=0$.

\section{Acknowledgments}

DJN thanks the kind hospitality of the Brown University Department of Chemistry. This project was partially supported by Fulbright COMEXUS Project NO. P000003405.

\appendix

\section{The classical period on the vicinity of the critical energy}
\label{FormulaLogaritmica}

The classical period given by (\ref{Tclasico}) can be conveniently written in terms of the turning
points as 
\begin{equation}
\label{periodo2}
T=\sqrt{\frac{2}{b}}\int_{x_1}^{x_2}\frac{dx}{\sqrt{(x-x_1)(x_2-x)(x+x_1)(x+x_2)}}.
\end{equation}
where $x_1,x_2$ are given by
\begin{equation}
\label{turningpoints}
x_{2,1}= \frac{\sqrt{-a\pm a\sqrt{1+ 4 b E/a^2}}}{\sqrt{2b}}.
\end{equation}

The indefinite integral has the following form
\begin{eqnarray}
\label{indefinida}
I(x,x_1,x_2)&\equiv&\int \frac{dx}{\sqrt{(x-x_1)(x_2-x)(x+x_1)(x+x_2)}} \nonumber \\
&=& \frac{2iF\left(\arcsin\left( \sqrt{\frac{(x_1+x_2)(x_2-x)}{2 x_2(x_1-x)}}\right)\mbox{\Large{$|$}} \frac{4 x_1 x_2}{(x_1+x_2)^2} \right) }{(x_1+x_2)},
\end{eqnarray}
where $F(\phi|u)$ is a incomplete elliptical integral of first kind. It is clear that (\ref{indefinida}) vanishes when evaluating in $x_2$. Before evaluating (\ref{indefinida}) in $x_1$, we should explore the behavior of the arcsin function in the limit $x\to x_1$
$$
\lim_{x\rightarrow x_1}\arcsin\left(\sqrt{\frac{(x_1+x_2)(x_2-x)}{2 x_2(x_1-x)}}\right)=\lim_{x\rightarrow x_1}\arcsin\left(i \sqrt{\frac{x_2^2-x_1^2}{2x_2}}\frac{1}{\sqrt{x-x_1}}\right)
$$
$$
\approx \frac{i}{2}\left(\log \frac{2(x_2^2-x_1^2)}{x_2}-\log (x-x_1) \right)\,,
$$
taking only the leading term of the series expansion.
By substituting this expression in the indefinite integral (\ref{indefinida}) we obtain
$$
\lim_{x\rightarrow x_1} I(x,x_1,x_2)=\frac{-2}{x_1+x_2}K\left(1-\frac{4 x_1 x_2}{(x_1+x_2)^2}\right),
$$
 where $K(u)$ is the complete elliptic integral of first kind.
Substituting the limit in (\ref{periodo2}) we obtain the classical period in terms of the turning points
$$
T=\sqrt{\frac{2}{b}}\left(\lim_{x\rightarrow x_2} I(x,x_1,x_2)-\lim_{x\rightarrow x_1} I(x,x_1,x_2)\right)$$
$$
=
\sqrt{\frac{2}{b}}\frac{2}{x_1+x_2}K\left(1-\frac{4 x_1 x_2}{(x_1+x_2)^2}\right)\,.
$$
So far, $x_1$ and $x_2$ have been treated as arbitrary only depending on the energy (\ref{turningpoints}). However we are interested in the turning points (\ref{turningpoints}) when the energy approaches the critical $E\to0$. In this limit case, we can use the approximation $$\sqrt{1+\frac{4bE}{a^2}}\approx1+\frac{2bE}{a^2}$$
to obtain the leading term of the turning points $x_2\rightarrow \sqrt{\frac{-a}{b}}$ and $\lim_{E\rightarrow 0^-}x_1=\sqrt{\frac{|E|}{-a}}. $ For this turning points close to the critical energy, the period is
$$
 \lim_{E\rightarrow 0^-}T=2\sqrt{\frac{2}{-a}}K\left(1-\frac{4 \sqrt{b |E|}}{|a|}\right).
$$

Finally, considering the asymptotic behavior of the complete elliptical integral
$$
\lim_{\epsilon\rightarrow 0}K(1-\epsilon)=\frac{1}{2}(4 \log 2-\log\epsilon), 
$$    
we find that the classical period behaves as
$$
\lim_{E\rightarrow 0^-}T=-\sqrt{\frac{2}{-a}}\left(\log \left[\frac{\sqrt{b}}{-4 a}\right]+\frac{1}{2}\log|E|\right).
$$
which in turns can be writen in terms of the Lyapunov exponent $\lambda=\sqrt{-2a}
$. Finally, by recalling that for $E<0$ the total (irrespective of the parity) density of states is $2\pi\hbar \rho(E)=2 T$, we obtain Eq.(\ref{linearrelation}) of the main text.

\section{The Lyapunov exponent}
\label{LyapunovAp}

In this appendix we discuss in detail how to obtain the value of the Lyapunov exponent corresponding to the stationary point ($x=0$,$p=0$) of the phase-space corresponding to the the quartic oscillator (\ref{Ham}). The canonical Hamilton equations read
\begin{eqnarray}
\label{HamEquations}
    \Dot{x} &=& \frac{\partial H}{\partial p} = p
    \nonumber \\
    \Dot{p} &=& - \frac{\partial H}{\partial x} = -4bx^{3} - 2ax
\end{eqnarray}
Therefore we can write
\begin{eqnarray} 
(\dot x, \dot p ) &=& F(x,p) 
\end{eqnarray}
with
\begin{equation}
\label{Fxp}
F(x , p) =  (p , -4bx^{3} - 2ax).
\end{equation}
A stationary stationary point satisfies $F(x_{0} , p_{0}) = (0,0)$, which is equivalent to $x(t) = x(t_{0})=x_0$ and $p(t) = p(t_{0})=p_0$. In order to find the Lyapunov exponent associated to $x_{0}$, let us build the Jacobian matrix $A = D_{x} F(x , p)$ and evaluate it at the stationary point  \cite{PilatowskyChavez}

\begin{equation}
\label{matriz}
A = D_{x} F(x , p) \arrowvert_{(x , p) = (0 , 0)} =
\left( 
\begin{array}{cc}

0 & 1\\

\\

-2a & 0

\end{array}
\right)
\end{equation}
The eigenvalues of (\ref{matriz}) are $\lambda = \pm \sqrt{-2a}
\,$. The Lyapunov exponent corresponds to the maximum real part of the eigenvalues. Since $a=-10$ in our case of interest, the Lyapunov exponent is the positive eigenvalue.



\bibliography{Qdots2020}

\end{document}